\documentclass[preprint2]{aastex}
\usepackage{natbib}

\shorttitle{Compact radio sources in the ONC}
\shortauthors{Forbrich et al.}

\begin{document}


\title{The Population of Compact Radio Sources in the Orion Nebula Cluster}




\author{J. Forbrich\altaffilmark{1}}
\affil{University of Vienna, Department of Astrophysics, T\"urkenschanzstr. 17, 1180 Vienna, Austria}
\email{jan.forbrich@univie.ac.at}

\author{V. M. Rivilla}
\affil{Osservatorio Astrofisico di Arcetri, Largo Enrico Fermi, 5, I-50125, Firenze, Italy}

\author{K. M. Menten}
\affil{Max-Planck-Institut f\"ur Radioastronomie, Auf dem H\"ugel 69, D-53121 Bonn, Germany}

\author{M. J. Reid}
\affil{Harvard-Smithsonian Center for Astrophysics, 60 Garden St MS 42, Cambridge, MA 02138, USA}

\author{C. J. Chandler, U. Rau, S. Bhatnagar}
\affil{National Radio Astronomy Observatory, P.O. Box O, Socorro, NM 87801, USA}

\author{S. J. Wolk}
\affil{Harvard-Smithsonian Center for Astrophysics, 60 Garden St MS 42, Cambridge, MA 02138, USA}

\and

\author{S. Meingast}
\affil{University of Vienna, Department of Astrophysics, T\"urkenschanzstr. 17, 1180 Vienna, Austria}


\altaffiltext{1}{also at: Harvard-Smithsonian Center for Astrophysics, 60 Garden St MS 72, Cambridge, MA 02138, USA}


\begin{abstract}
We present a deep centimeter-wavelength catalog of the Orion Nebula Cluster (ONC), based on a 30~h single-pointing observation with the Karl G. Jansky Very Large Array in its high-resolution A-configuration using two 1~GHz bands centered at 4.7~GHz and 7.3~GHz.  A total of 556 compact sources were detected in a map with a nominal rms noise of 3~$\mu$Jy\,bm$^{-1}$, limited by complex source structure and the primary beam response.   Compared to previous catalogs, our detections increase the sample of known compact radio sources in the ONC by more than a factor of seven.  The new data show complex emission on a wide range of spatial scales. Following a preliminary correction for the wideband primary-beam response, we determine radio spectral indices for 170 sources whose index uncertainties are less than $\pm0.5$.  We compare the radio to the X-ray and near-infrared point-source populations, noting similarities and differences. 
\end{abstract}





\section{Introduction}

\begin{figure*}
\includegraphics[width=0.49\linewidth]{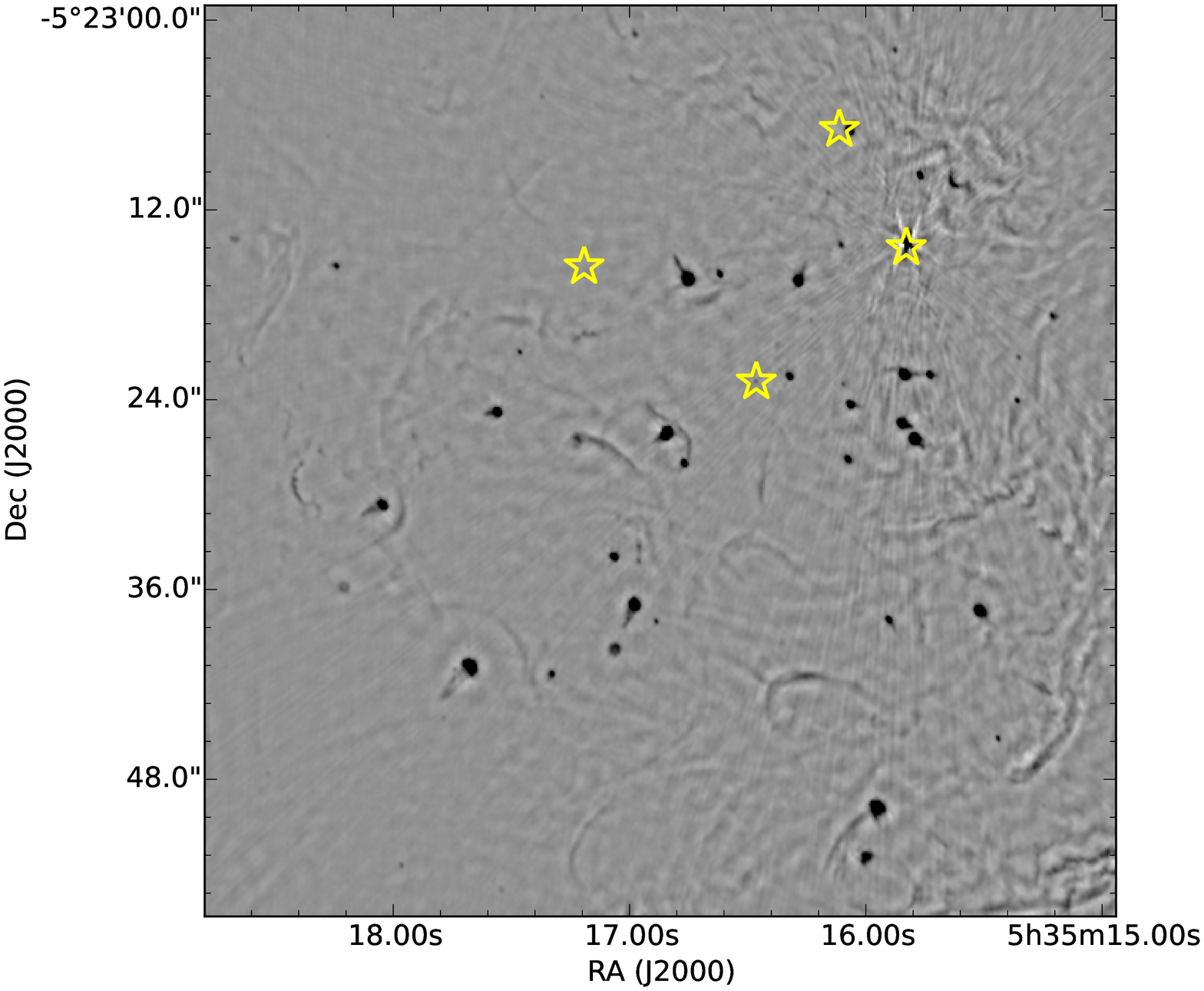}
\includegraphics[width=0.49\linewidth]{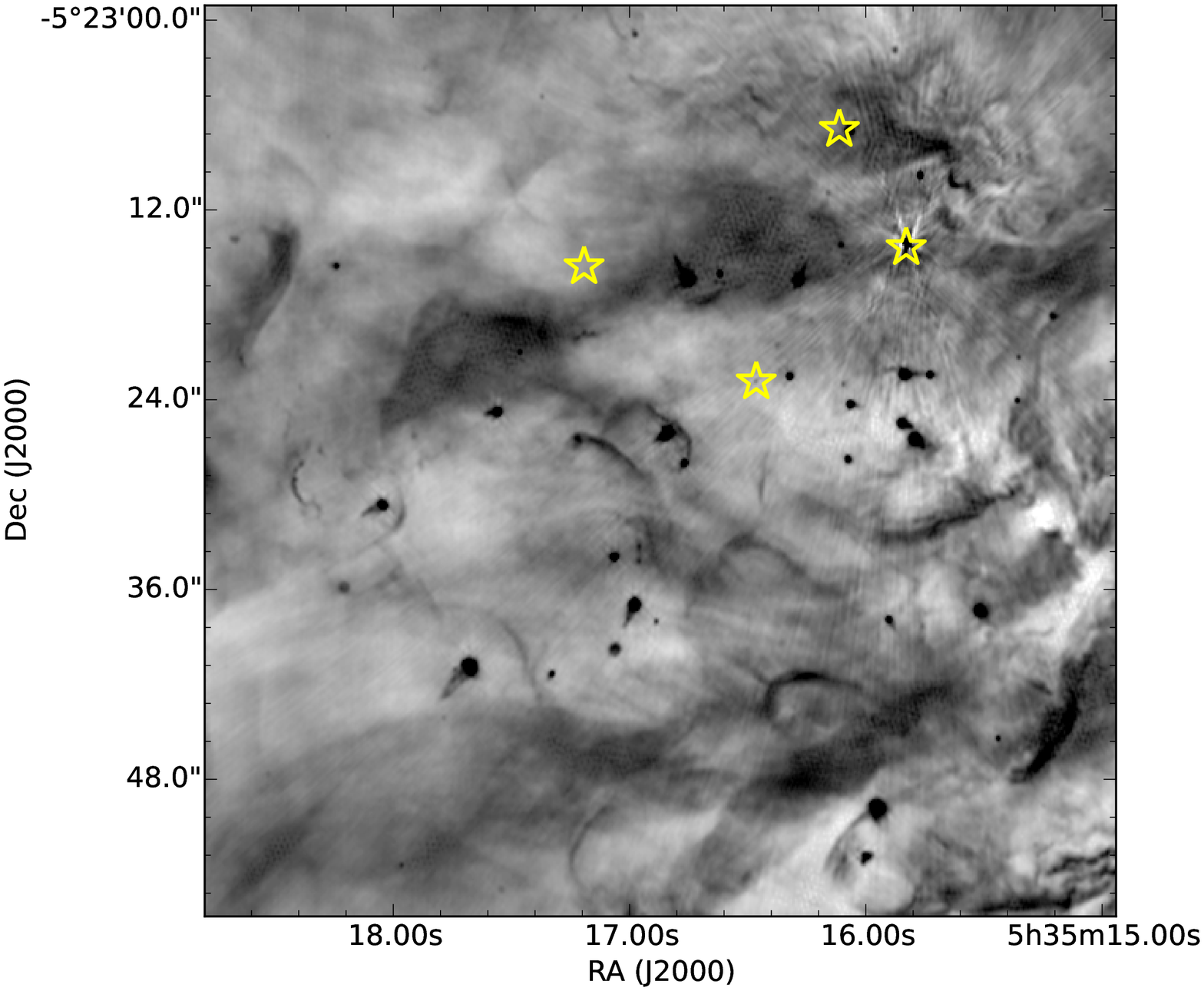}
\caption{Inner area of the field of view and the cluster, with (left) and without (right) additional spatial filtering (100k$\lambda$). These images show the full wideband data with a reference frequency of 6.1~GHz. The color scale is the same for both frames. It has been selected to highlight weaker features, ranging from --0.3 to +0.4 mJy\,bm$^{-1}$ in linear scaling. The locations of the main Trapezium stars $\theta^1$ Ori A--D are marked with star symbols. Many different types of sources are apparent in both images.\label{fig_inner}}
\end{figure*}

The Orion Nebula Cluster (ONC) has long been a benchmark for centimeter wavelength studies of young stars. It is here that the first compact centimeter radio counterparts of young stars were discovered using radio interferometry with the National Radio Astronomy Observatory's Very Large Array \citep{mor82,mor83}, prior to its recent upgrade to the Karl G. Jansky Very Large Array (VLA). Subsequent studies uncovered an entire `zoo' of compact radio sources in the ONC \citep{gar87,chu87,fel93b,hen01}. Prior to the VLA upgrade, centimeter radio studies of the ONC culminated in the deep high-resolution continuum survey by \citet{zap04}. They detected 77 radio sources in an image with an rms noise level of 30~$\mu$Jy\,bm$^{-1}$ (1$\sigma$), many of them variable. The first observations using continuum Very Long Baseline Interferometry (VLBI) toward sources in this region were carried out by \citet{fel89}, targeting $\theta^1$ Ori A. Subsequently, \citet{men07} obtained astrometric continuum VLBI observations of four ONC stars to obtain a parallax distance of $d=414\pm7$~pc (see also \citealp{kim08}). \citet{kou14} used the upgraded VLA to obtain a wide-area census of radio sources in the Orion region down to an rms noise level of 60~$\mu$Jy\,bm$^{-1}$ (1$\sigma$).  More recently, we presented a new assessment of Q and Ka band radio variability in the Orion BN/KL region \citep{riv15}.

Regarding radio emission surveys of the ONC, several issues are important. Firstly, the entire Orion Nebula is a very bright and extended source of about 400~Jy at centimeter wavelengths. Embedded within this nebula, the ONC contains compact sources with flux densities of less than 0.1~mJy, which corresponds to a dynamic range of up to $10^6$.  It is here that interferometric imaging with its spatial filtering is an advantage, because it reduces the imaging dynamic range requirement (the entire radio flux, of course, still causes elevated system temperatures).

Secondly, there is a wide, continuous range of source size scales, from very extended emission to very compact sources, and it can be unclear how to categorize sources. Thermal free-free emission from ionized material is observed on many scales (as is dust emission), while nonthermal emission from stellar coronal activity (e.g., \citealp{dul85,gue02}) is very compact. Mildly or highly relativistic electrons gyrate in magnetic fields, emitting gyrosynchrotron or synchrotron emission. Observational means of distinguishing non-thermal from thermal emission involve, in approximate order of reliability, polarization (circular for gyrosynchrotron emission, linear for synchrotron emission), spectral indices ($S_\nu\propto\nu^\alpha$ with spectra rising or flat for thermal and falling for non-thermal sources), and the rapid variability usually associated with non-thermal sources.  All of these criteria require high signal to noise ratios (SNRs), and thus for weak sources it may be impossible to determine the emission mechanism in practice. 

Thirdly, while centimeter radio emission is insensitive to interstellar extinction, in contrast to infrared and X-ray emission, plasma can easily become optically thick, rendering sources behind it unobservable. Such foreground plasma may not even emit radiation at the observed frequency (e.g., \citealp{gue02}), or its emission may be largely filtered out by an interferometer if occurring on larger spatial scales. A radio non-detection thus does not necessarily mean that a star is not emitting at radio wavelengths at levels that would otherwise be detectable.

\begin{figure}
\includegraphics*[width=\linewidth]{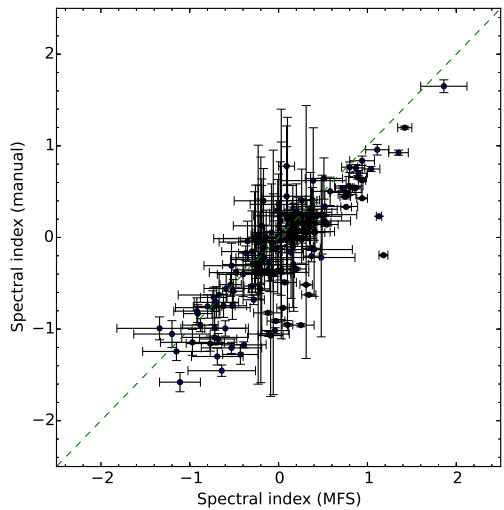}
\caption{Comparison of spectral indices as derived by MFS and in a manual 2-band calculation (see text). Note that the error bars are not directly comparable since they are based on the MFS residual map and simple error propagation for the flux densities in two bands. All points would lie on the dashed line if the spectral indices were identical in both methods.\label{fig_compalpha}}
\end{figure}

Young Stellar Objects (YSOs) show both nonthermal radio and X-ray emission. The two emission processes are thought to be due to coronal activity in these sources (e.g., \citealp{fem99}); they may also show thermal radio emission. Based on deep observations of YSOs, most notably in the Chandra Orion Ultra-deep Project (COUP), the ONC has become one of the best-characterized young clusters in X-rays (\citealp{get05} and references therein). In 2013, we investigated the relation between X-ray and radio emission in the ONC, based on the most sensitive radio data prior to the VLA upgrade and non-simultaneous X-ray data \citep{fow13}. A review of simultaneous X-ray and radio observations of YSOs has been published by \citet{for11,for11cs}.

We here present a deep radio census of point-like sources in the ONC, using a single deep pointing of about 30~hours with the upgraded VLA, obtained in five epochs spanning six days. The nominal sensitivity of the concatenated data is 3~$\mu$Jy\,bm$^{-1}$ (1$\sigma$) at the center of the primary beam, elevated by contributions from complex emission structure and the primary beam response (see below).  This is up to 10 times better than the most sensitive previous study \citep{zap04}.  Our observations were obtained with largely simultaneous X-ray coverage, using the {\it Chandra} X-ray Observatory, the main driver being the study of simultaneous X-ray and radio variability. 

This paper discusses the nature of X-ray and near-infrared counterparts to the radio sources. In Section~\ref{sec:obs} 
we describe the VLA observations, followed by a discussion of the data reduction in Section~\ref{sec:data}.  This includes the definition of the compact source catalog and its interpretation in the context of X-ray and near-infrared studies. After discussing notable individual sources, we close with a summary in Section~\ref{sec:sum}.

In follow-up papers, we will address the following issues. Since the increase in radio continuum bandwidth means that the assumption of monochromaticity has to be abandoned, new imaging techniques are required \citep{bha13}, with implications for all of the above-mentioned means of distinguishing thermal and nonthermal emission. Note, for example, that in our experiment the size of the primary beam changes by a factor of two over the sampled frequency range.  In a follow-up paper we will thus assess the effect of wideband imaging on polarization measurements and its implications for emission mechanisms. We will also discuss simultaneous radio and X-ray variability at high time resolution in more detail later. Finally, even though the characterization of radio counterparts of YSOs was the main motivation for this experiment, the data also reveal exquisite details of extended emission, ranging from proplyds and Herbig-Haro objects to even more extended emission from the Orion Nebula itself.

\begin{figure}
\includegraphics*[width=\linewidth,bb= -44 96 575 659]{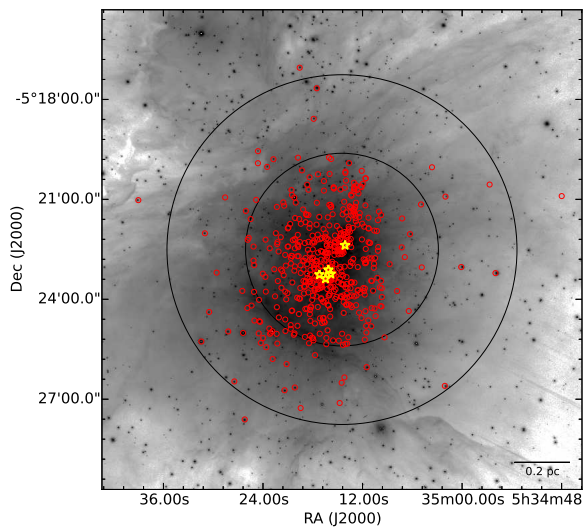}
\caption{Distribution of the 556 point-like sources, overlaid on the VISION-$K_S$ image from \citet{mei16}. For orientation, the locations of the Trapezium ($\theta^1$ Ori A--D) and the BN object are marked with yellow star symbols. The circles indicate the smallest and largest HPBW primary beam, and the scale bar indicates the physical scale at a disctance of 414~pc.\label{fig_psc}}
\end{figure}

\begin{figure}
\includegraphics*[width=\linewidth]{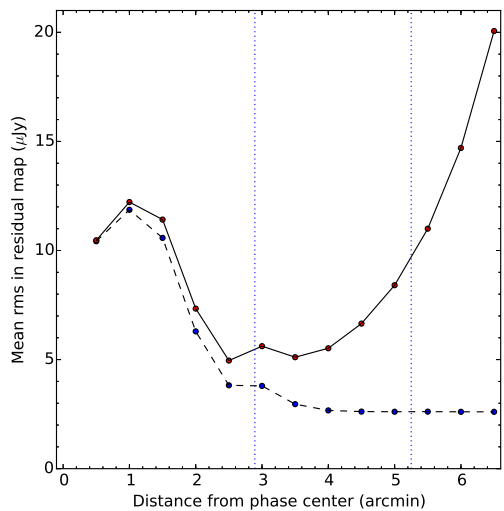}
\caption{Rms noise in the residual image of the spatially filtered ($(u,v)>100$k$\lambda$) dataset. The upper (red) points, connected by a continous line, were corrected for the wideband primary beam response. The uncorrected (blue) points, connected by a dashed line, are shown for comparison. The dotted vertical lines indicate the radii of the smallest and largest half-power primary beam. \label{fig_rms_radial}}
\end{figure}

\section{Observations\label{sec:obs}}

The observations were carried out with the Karl G. Jansky Very Large Array (VLA) of the National Radio Astronomy Observatory\footnote{The National Radio Astronomy Observatory is a facility of the National Science Foundation operated under cooperative agreement by Associated Universities, Inc.} on 2012 September 30, October 2, 3, 4, and 5 under project code SD630. The phase center was at (RA,Dec) = $05^h35^m14.479^s,-5^\circ22'30.57''$ (J2000); the durations of these individual epochs were generally 7.5~h, except for epoch 3 with about 3~h, and epoch 4, with about 5~h, resulting in a total of about 30~h.  During the first two epochs, the array was being re-configured to its high-resolution A-configuration from the BnA-configuration.  For the other three epochs, the array was in its A-configuration. This configuration was chosen to maximize spatial filtering and optimize the detection of point-like sources. 

Data were taken using the VLA's C-band (4-8 GHz) receivers in full polarization mode, with two 1 GHz basebands centered at 4.736 and 7.336 GHz to provide a good baseline for source spectral index determination.  Each baseband was split into eight subbands of 128 MHz each, which were in turn divided into 64$\times$2~MHz spectral channels. 3C~48 served as the primary flux density calibrator and J0541-0541 was observed in order to monitor the complex gains.  Generally, after five minutes on-target (ONC), we switched to the gain calibrator, ensuring very stable amplitude and phase calibration.

Apart from the first epoch, the field was simultaneously observed with the {\it Chandra} X-ray Observatory. Mostly of interest for variability information, these data will be presented as part of a follow-up paper.  Generally, the COUP survey -- based on significantly longer and thus more sensitive observations -- has resulted in the benchmark X-ray catalog of this region. In order to identify X-ray counterparts to radio sources, we thus primarily use the COUP catalog \citep{get05}. To identify infrared counterparts to the radio sources, we make use of a new near-infrared ($JHK_S$) point-source catalog of Orion~A \citep{mei16}.

\begin{figure*}
\includegraphics[width=0.49\linewidth]{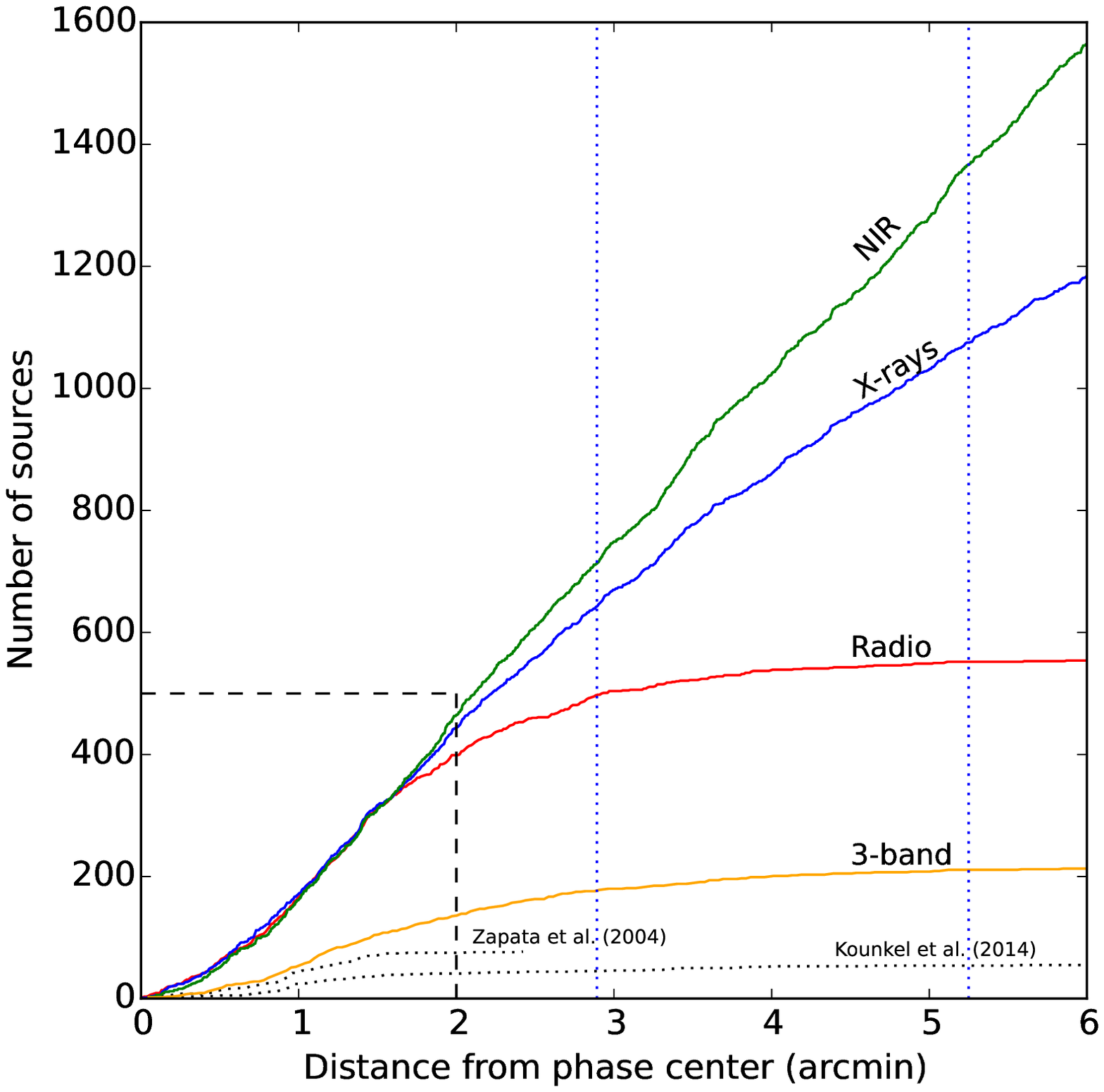}
\includegraphics[width=0.49\linewidth]{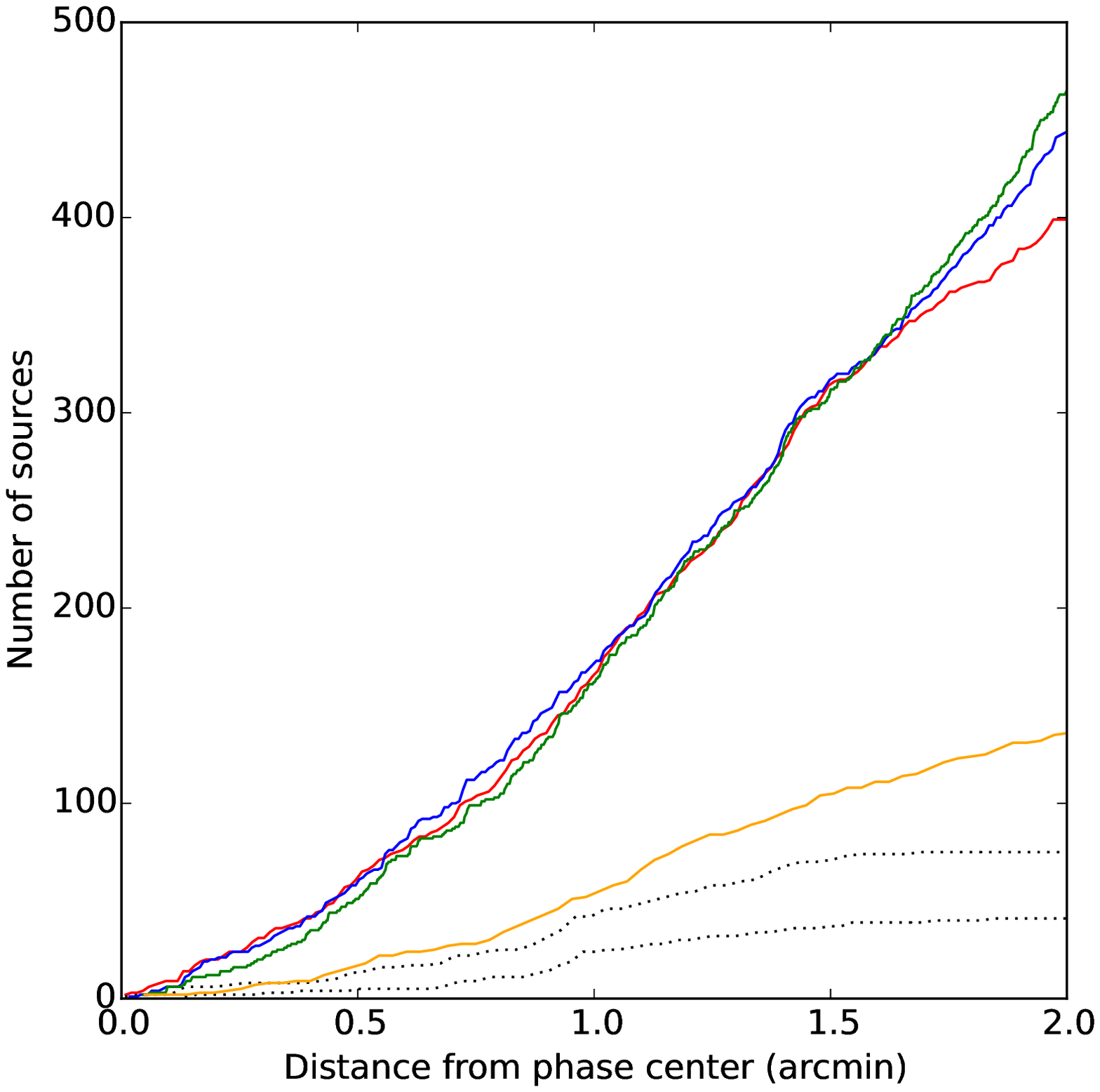}
\caption{Left: Cumulative source distribution as a function of radial distance. Right: Detailed view of the inner beam. In both panels the red, blue, and green lines indicate the radio, X-ray, and infrared populations, respectively, while the orange line denotes the population of 3-band detections. The black dotted lines indicate the previously known radio sources from \citet{zap04} and \citet{kou14}. In the left panel, the vertical blue dotted lines delineate the radii of the smallest and largest half-power primary beam, and the dashed lines indicate the data range shown in detail in the right-hand panel.\label{fig_cumdist1}}
\end{figure*}

\section{Data reduction and analysis\label{sec:data}}

The data were reduced using the VLA pipeline \footnote{\url{http://science.nrao.edu/facilities/vla/data-processing/pipeline}}, which involves manual interaction for `flagging' (i.e., the identification of faulty data not to be used further), particularly of radio frequency interference. The pipeline uses the Common Astronomy Software Applications (CASA) environment, specifically the CASA~4.2.2 release. Imaging was carried out within CASA and some parts of the subsequent analysis were run in the
Astronomical Image Processing System (AIPS).

The five epochs were reduced separately. Initially, to better cope with the amount of data, we conducted a series of tests as to how much time 
and spectral averaging could be done with acceptable losses across the wide-field we imaged.  We decided to apply no additional temporal averaging (beyond the 1-second correlator integrations), but we averaged the 64 spectral channels down to 4 channels, leading to a loss in peak flux density for a source in the outer beam of $<6$\% by bandwidth smearing (chromatic aberration). The test source in this case was KM~Ori (source 2) at a distance of 4$\farcm$7 from the phase center where the wideband primary beam attenuation amounts to a factor of 3.1 relative to the phase center. Due to the rapid switching to the phase calibrator, phase-only self-calibration on the data did not further improve the quality of the phase calibration. 

After calibration, all (u,v)-data were imaged jointly using the `clean' task in CASA.  A multi-frequency synthesis (MFS) approach, which models the frequency dependence of the source emission as a Taylor series during the deconvolution of the point spread function \citep{rau11} enabled maximum continuum sensitivity.  A Taylor series with nterms=2 in 'clean' was used.  The output of this algorithm are coefficients of the Taylor expansion and their residual images following the deconvolution, which can subsequently be used to compute the total intensity image at the reference frequency (in this case, 6.1 GHz), and the spectral index of the equivalent power law of the emission.  The errors for the spectral index map are estimated by propagating appropriately the error terms derived from the residual images.

We generated a Stokes-$I$ image of 8192$\times$8192 pixels of 0$\farcs$1, covering a field of view of  13$\farcm$6 on a side.  This setup was mainly chosen to cover the half-power beamwidth at the low-frequency end of the bandwidth at 4.3~GHz ($\sim10\farcm5$ HPBW), and the image thus covers more than the half-power beam for the high-frequency end of the bandwidth at 7.8~GHz ($\sim5\farcm8$ HPBW). Briggs weighting of the $(u,v)$ data with a `robust' parameter of zero was used, as an optimal balance between resolution and sensitivity. Images with both the full $(u,v)$-data and with additional spatial filtering, with a $(u,v)$ range set to $>100$~k$\lambda$, were produced. We found that this low spatial-frequency cutoff constitutes a good compromise between point source sensitivity and spatial filtering. To assess the spectral indices in the field, we used the CASA task `widebandpbcor' (in CASA 4.4) to obtain images corrected to zero order for the wideband primary beam response. With this task, a correction factor was calculated for every spectral window. Using `nterms=2', this task also produces a map of corrected spectral indices.

In Figure~\ref{fig_inner}, we show VLA images of the innermost region of the cluster in the vicinity of the Trapezium.  The right panel shows the image based on the entire $(u,v)$-data, while the left panel shows the image created with the limited $(u,v)$ range, which filters structures larger than about $\sim$2$''$. This is still larger than the synthesized beam size of $0\farcs30\times0\farcs19$ (FWHM) at PA 30$^\circ$. A visual inspection of these VLA images of the inner ONC with optical images (e.g., from the Hubble Space Telescope), we concur with \citet{fel93} who observed a good correspondence of radio structures with H$_\alpha$ emission on all scale sizes.

\subsection{Assessment of spectral indices}

Our assessment of the spectral indices is based on the spectral index map produced by the tasks `clean' and `widebandpbcor'. For resolved structure, such as for closely spaced sources, the MFS approach is able to extract spectral index information at the resolution of the MFS map, which is usually sharper than just the low frequency map. Also, complicated structure benefits from using all the data simultaneously during image reconstruction to constrain both structure and spectrum at the same time.  With the traditional separate approach, due to the non-uniqueness of deconvolution results, one could end up with two separate reconstructions that look fine on their own but which are inconsistent when the spectrum is computed (for a more detailed discussion, see \citealp{rau11}). 

For comparison, we have also imaged the two 1~GHz bands separately, to then obtain spectral indices for unresolved and isolated sources that are detected in both bands, after convolving the two images to the same beam size (given by the low-frequency band) and applying individual primary beam corrections from `widebandpbcor'. We then used the AIPS task JMFIT to obtain background-subtracted flux densities in both bands to calculate the corresponding spectral indices. We compared this to simply integrating the flux densities in small boxes around each source position, but this method fails when there are different levels of extended emission. Also, using JMFIT for this comparison fails when a source cannot be described by a Gaussian function, for example when it is resolved with a complex shape. 

As shown in Figure~\ref{fig_compalpha}, we found good general agreement between the two sets of spectral indices for 157 sources where the spectral index error determined from the MFS images is $\leq0.5$ and JMFIT peak flux densities could be determined. A few clear outliers highlight difficulties with resolved sources in this simple test. Overall, when using the MFS spectral indices instead of the traditional approach, a larger number of spectral index measurements is available below a given cutoff in the nominal spectral index errors \citep{rau11}.

\subsection{Point source detection}

It is obvious from Figure~\ref{fig_inner} that, even with the spatial filtering afforded by a $(u,v)$ cutoff at $>100$~k$\lambda$ when imaging, bona-fide point sources that are seen in projection against extended structure will be distorted.  Additional complications can arise from imaging artifacts due to temporal variability, effectively limiting the sensitivity and dynamic range in portions of the image. An example for such artifacts can be seen in the vicinity of $\theta^1$ Ori A, the westernmost Trapezium star in the top right corner of Figure~\ref{fig_inner}. Also, some apparent point sources, most likely representing ionized envelopes of young stars, can appear marginally resolved. Since our primary objective here is the detection of radio counterparts of stars, we focus on the detection of pointlike compact sources.

Due to the spatial complexity of the data, we found that no automated point-source detection scheme resulted in a reliable list of detections. Up to 50\% of the sources meeting formal selection criteria based on signal-to-noise turned out to be spurious because of artifacts and sidelobes. With a high degree of manual intervention required, we decided to perform visual source detection first, carried out in several independent runs on all individual epochs, and then in the concatenated data.  A combined total of almost 700 candidate sources was identified and visually checked in the deep image obtained from the concatenated $(u,v)$-data. As preliminary indicators to assess these candidates, we also considered the local SNR on scales of $\sim$2$''$ and $\sim$4$''$ (9$\times$9 and 19$\times$19 pixels).  Given the spatial complexity of the radio emission, there is, however, no generally useful cutoff criterion in the local SNR, but 98\% of our sources have a local SNR$>$3. In a second step, we used the AIPS task JMFIT to obtain a Gaussian fit to each source, which was then used to enforce a consistent signal-to-noise cutoff of SNR$>$5 in the concatenated data, where the task mainly uses the global rms to calculate SNR. We report the peak flux density and its error, both corrected for the primary beam response. 

Our catalog of detections above SNR$>$5 contains 556 sources and is listed in Table~\ref{tbl-1}, their locations are shown in Figure~\ref{fig_psc}. The 556 sources have median position errors, as estimated by JMFIT, of 6~mas in RA and 7~mas in Dec, with maximum errors of $0\farcs1$. These errors simply reflect the S/N ratio, and since not all of our sources are true point sources, this error reflects both the fit and potentially also source structure. We can estimate the absolute astrometric accuracy of our observations by assessing the position of our source 8, which is bright and which has been identified as likely extragalactic \citep{get05b}. The source is at a distance of $2\farcm45$ from the phase center, and it has a peak flux density of 3.6~mJy\,bm$^{-1}$. Among the five individual epochs, the position varies with a 1$\sigma$ rms of 25~mas, consisting of an rms of 24~mas in RA and 7~mas in Declination, which is reflecting the asymmetric synthesized beam size and systematic errors. Based on these considerations, we thus estimate an overall absolute astrometric accuracy of 20 to 30~mas.

The presence of emission on many different spatial scales in the inner cluster means that it is difficult to assess the completeness of our catalog.  The main challenge comes from the fact that the nominal sensitivity is only reached in the outer regions of the cluster, but not in the innermost regions, due to the presence of artifacts and extended structure.  As a consequence, we have a consistent completeness limit in the outer regions that are only affected by the shape of the wideband primary beam, but not in the innermost regions of the cluster. 

In order to attempt to quantify the variation in catalog completeness due to complex emission in the inner cluster, as well as the decrease in sensitivity in the outer beam due to the primary beam response, we have calculated annular averages of the noise in the residual maps that were corrected for the primary beam response. In Figure~\ref{fig_rms_radial}, the resulting rms noise as a function of distance from the phase center is shown, both with and without a correction for the primary beam response. It becomes clear that the rms noise in the inner cluster is roughly 2--3 times as high in the area surrounding the phase center than further out, reaching $\sim12\mu$Jy\,bm$^{-1}$. The lowest rms noise of $\sim5\mu$Jy\,bm$^{-1}$ is then reached at about the radius of the HPBW of the highest-frequency primary beam, and it then rises steadily with increasing radius because of the correction for the wideband primary beam response. Nominally, however, without the primary beam correction, the rms noise level should be as low as 3~$\mu$Jy\,bm$^{-1}$ in the entire map. Note that the rms noise estimated in Figure~\ref{fig_rms_radial} does not necessarily correspond to the rms noise reported by JMFIT for an individual source.

\subsection{Previously known sources}

A different way of assessing the completeness is to compare the radio sample against previous radio catalogs as well as against the X-ray and near-infrared point source populations. To minimize confusion of closely spaced sources, we have opted for a consistent search radius of $0\farcs5$ in these searches for known counterparts in catalogs with well-matched spatial resolution. Only a few additional sources are found with larger search radii. 

Our catalog of 556 point-like radio sources constitutes the largest catalog of radio sources in the inner ONC.  Emission is detected from all 77 positions reported in the VLA study by \citet{zap04}, targeting the same area with the pre-upgrade VLA.  Additionally, most of the sources reported in the wide-area JVLA survey of \citet{kou14} were also detected in our experiment. They reported 55 sources within 6$'$ of our phase center at a quoted sensitivity of 60~$\mu$Jy (1$\sigma$). In this area, only four bright sources reported by \citet{kou14} remain undetected in our experiment, possibly indicating some highly variable sources. The brightest source is J053515.53--052248.3, listed by \citet{kou14} with a peak flux density at 4.5~GHz of 2.81~mJy.  At this position, our local 5$\sigma$ upper limit is 80~$\mu$Jy at a reference frequency of 6.1~GHz, which indicates variability by a factor of $\sim30$. Similarly, for J053504.55--052013.9, listed at 0.75~mJy by \citet{kou14}, we find an upper limit of 25~$\mu$Jy, and for J053521.66-052526.5, listed at 1.34~mJy, our upper limit is 30~$\mu$Jy. These cases indicate variability by a factor of at least $\sim30$ and $\sim40$, respectively, and are the only ones among these four with X-ray or infrared counterparts (within 0$\farcs$5 of the published position), the same criterion that we apply to our data (see below; but note that the aforementioned brightest source J053515.53--052248.3 has counterparts within 1$\farcs$0). The fourth example, J053517.32--052234.9, is less clear because it is 1$\farcs$1 (i.e., several synthesized beam sizes) away from a bright proplyd corresponding to our source 347 which itself does not seem to appear in the catalog of \citet{kou14}.

Two most obvious complementary datasets at other wavebands come from X-ray and near-infrared surveys. Here, we use the X-ray catalog of the Chandra Orion Ultra-deep Project (COUP), as reported by \citet{get05}, as well as the near-infrared data ($JHK_S$) from a new VISTA survey of Orion A, called VISION, reported by \citet{mei16}.  All of these datasets are expected to trace ONC members very well, missing only the most deeply embedded sources.  In the near-infrared, additional complications include confusion due to the bright extended nebula, a problem that does not occur in the X-ray observations, and the presence of unrelated foreground stars.

All but four of our compact radio sources lie within a distance of $5\farcm2$ of the phase center, spanning an area that corresponds approximately to the largest half-power primary beam in our VLA observations.  In the same area, where we find 552 compact radio sources, the COUP catalog lists 1067 X-ray sources, and VISION 1354 near-infrared sources.

\begin{figure}
\includegraphics[width=1.1\linewidth]{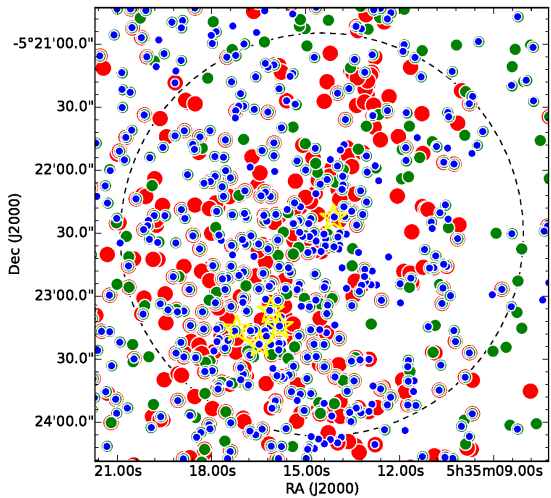}
\caption{Inner area of the cluster, with radio (red), infrared (green), and X-ray (blue) sources indicated. The symbol sizes have been chosen for clarity and do not correspond to the respective angular resolution. For orientation, the positions of $\theta^1$ Ori A--D and the BN object are marked with yellow star symbols, and the dashed circle indicates a radius of $1\farcm$6, as used for the Venn diagram in Figure~\ref{fig_vennXIR}. \label{fig_innerarea_XIR}}
\end{figure}

\subsection{Source distribution}

To study the radio source distribution in more detail, and to compare it with the distribution of X-ray and near-infrared sources, we plot the cumulative radial source counts in these three bands, as well as the distributions of sources with detections in all three of these bands in Figure~\ref{fig_cumdist1}.  Interestingly, the three distributions rise almost in unison with increasing distance from the phase center, up to a radius of about $1\farcm6$.  At this point, the radio source distribution starts to level off, at least partly due to the decreasing sensitivity caused by radially increasing attenuation due to the primary beam response.  The X-ray and infrared distributions continue to rise, with the X-ray distribution slowly leveling off at a radial distance of $\sim2'$. This leveling off is partly attributable to a drop in sensitivity away form the center of the field of view, and it partly seems to reflect actual structure of the cluster.

Quite surprisingly, however, the almost identical rise in the three distributions does not mean that the three source populations are identical. A plot of the inner cluster, with the three different source populations indicated, is shown in Figure~\ref{fig_innerarea_XIR}.  While there clearly are similarities, there are also significant differences. The region within the radius of $1\farcm6$ contains 329 radio, 330 X-ray, and 334 near-infrared sources. Only about half of the radio sources have X-ray or infrared counterparts, both of these about equally likely. Most of the radio sources with a counterpart in one of these two bands have counterparts in both bands. 

\begin{figure}
\includegraphics*[width=\linewidth]{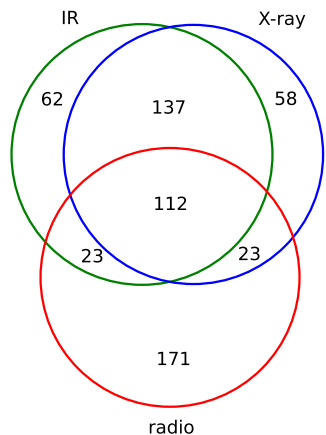}
\caption{Venn diagram of the source populations within $1\farcm$6 of the phase center, with area-proportional overlap. This area contains 329 radio, 330 X-ray, and 334 near-infrared sources. Note that this area does not contain the entire radio compact source catalog but is largely unaffected by the sensitivity drop toward the outer primary beam.\label{fig_vennXIR}}
\end{figure}

To better visualize these distinctions, we show a Venn diagram of the radio, X-ray, and near-infrared source populations in the inner r=$1\farcm6$ in Figure~\ref{fig_vennXIR}. Clearly, there are distinct populations in all three wavebands.  It may be no surprise to find that there are radio sources without counterparts at other wavelengths, because of the method of source detection.  We include all compact-looking sources, certainly including cases of thermal emission in shocks with no embedded sources that would generate infrared or X-ray emission. Sources with detections in at least two bands are most likely cluster members, particularly in this innermost area of the cluster, but a few of the weaker ones could be extragalactic background sources. X-ray and near-infrared sources are more likely to have multi-wavelength counterparts in the other bands than radio sources, possibly because they are more likely to be associated with young stars. In particular, the probability of an X-ray source having a near-infrared counterpart and vice versa is very high, 
at $\sim75$\%.

\begin{figure}
\centering
\includegraphics*[width=0.8\linewidth, bb= 13 36 600 530]{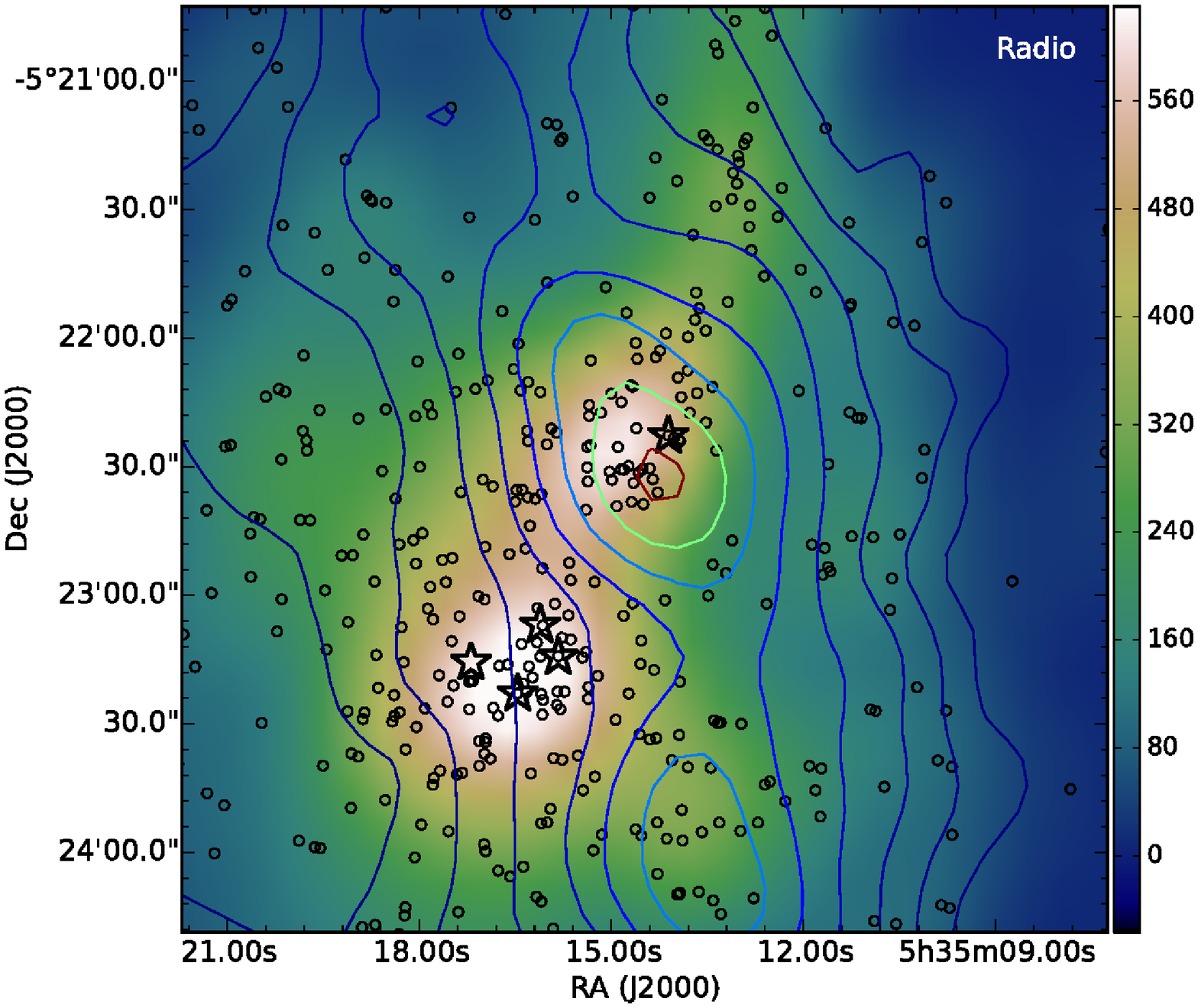}
\includegraphics*[width=0.8\linewidth, bb= 13 36 600 530]{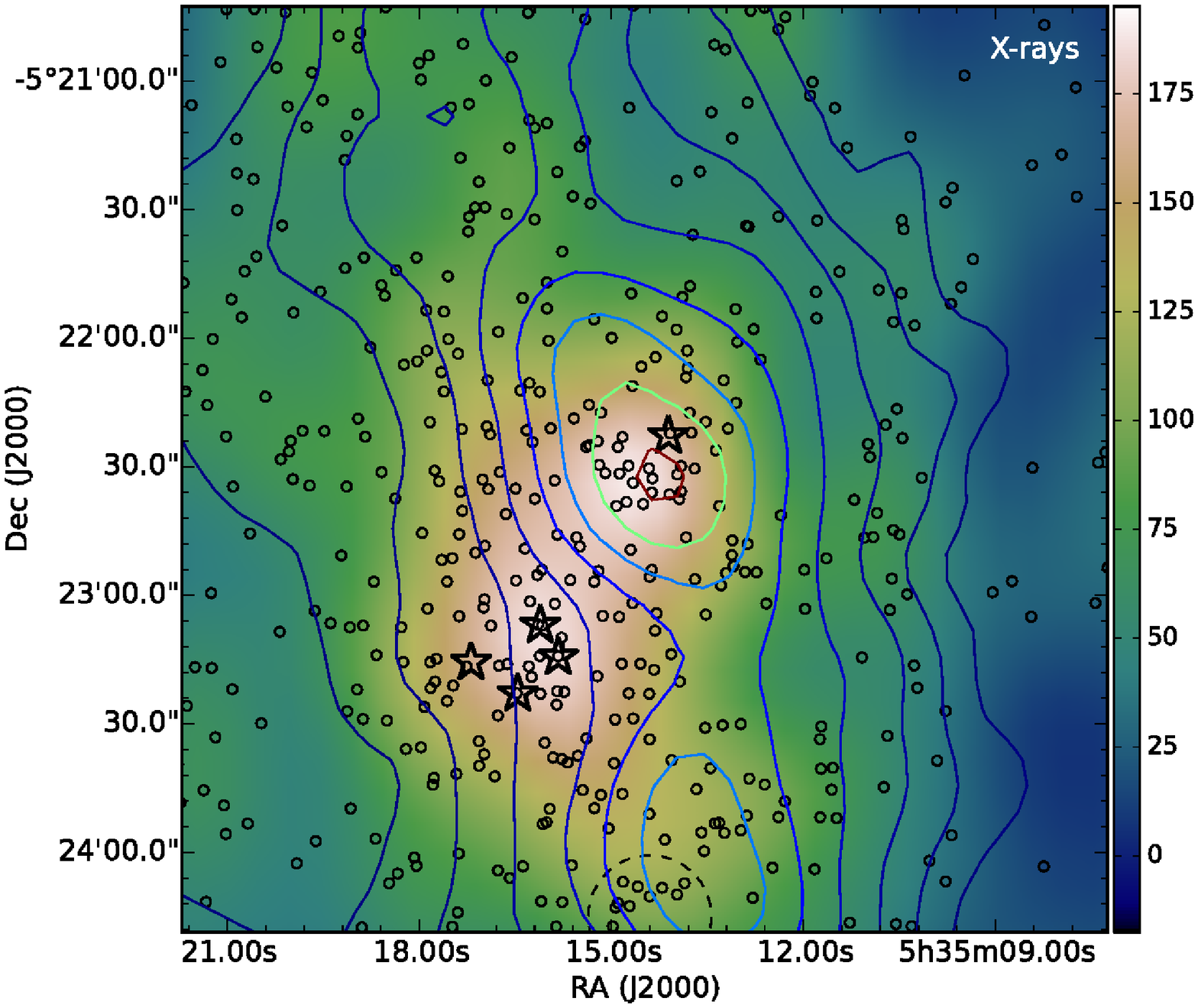}
\includegraphics*[width=0.8\linewidth, bb= 13 36 600 530]{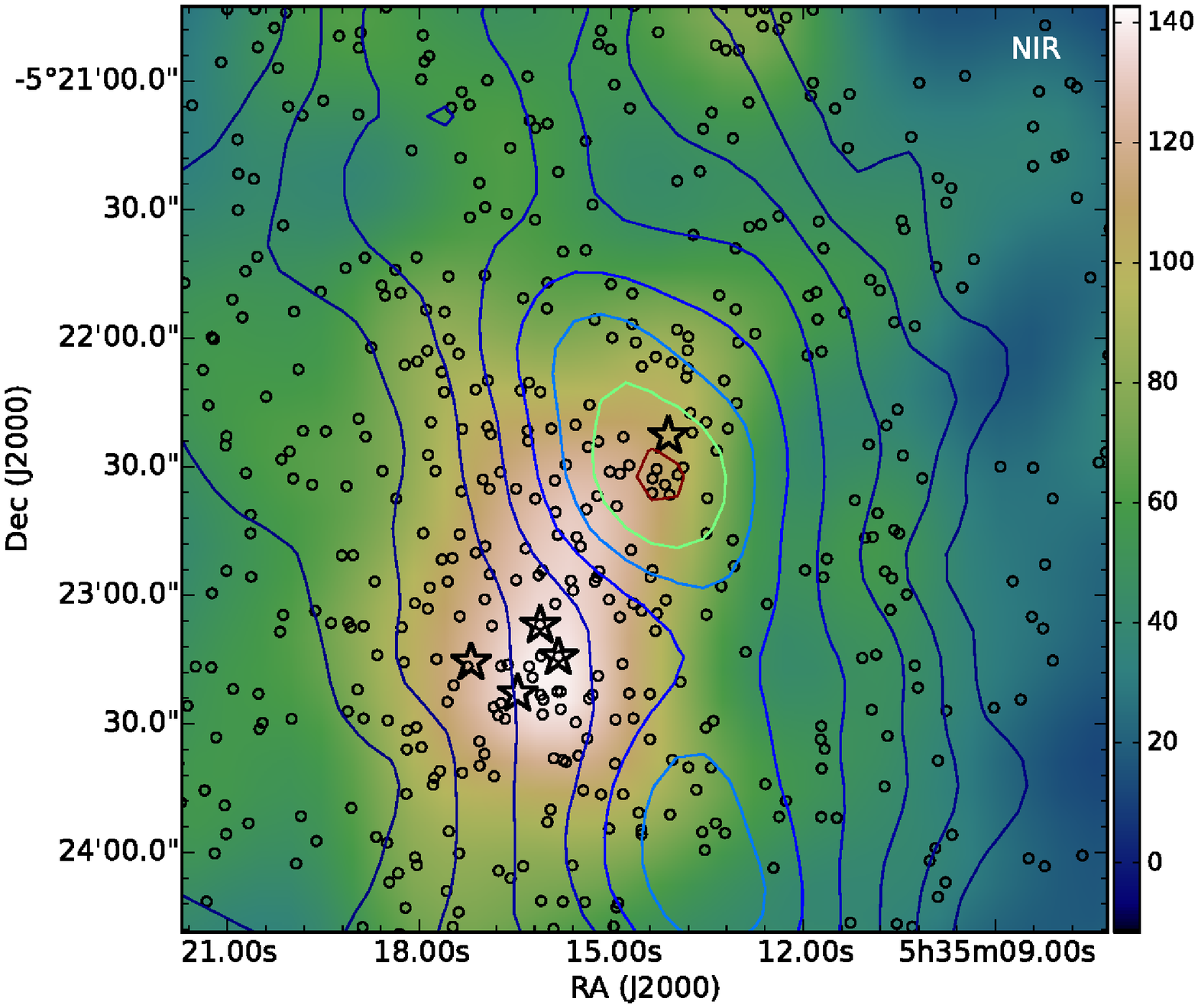}
\caption{Gaussian kernel density estimation (in color; bandwidth=15$''$) for the radio, X-ray, and infrared populations (upper, middle, and lower panels, respectively) of the inner ONC. Circles indicate the actual sources, and star symbols denote the locations of the Trapezium stars as well as BN. Contours indicate dust continuum emission as seen by SCUBA at 850~$\mu$m \citep{dif08}, with contours indicating factors of two in flux density. In the X-ray (middle) panel, a dashed circle at the Southern end of the plot indicates the position of OMC1-S, see text. \label{fig_spatialdist}}
\end{figure}

After assessing the radial properties of the source distributions, we now study these in more detail in two dimensions.  Figure~\ref{fig_spatialdist} shows the results of a Gaussian kernel density estimation, run on all three samples with the same bandwidth of 15$''$ to set the smoothing size scale. The similarity of the spatial source distributions at X-ray and near-infrared wavelengths are evident, while the radio source distribution is noticeably different.  In the radio, the BN/KL complex to the Northwest of the Trapezium clearly stands out, but is less obvious at X-ray and near-infrared wavelengths (see also \citealp{riv13a,kuh15}).  Overall, the radio source distribution is more elongated in the North-South direction than the other bands.  Generally, however, even the 2D distributions are similar enough to such a degree that the corresponding 1D profiles are almost identical.  Interestingly, we did not detect many sources toward another peak in the stellar density, OMC1-S.  This dense region in OMC harbors very extincted COUP X-ray sources that power several molecular outflows \citep{riv13b}.  These sources remain largely undetected in the VISION near-infrared data, as might be expected for deeply embedded sources, and they are also barely detected in our radio data (see Figure~\ref{fig_spatialdist}). Either these sources have intrinsically lower radio emission, or their emission is absorbed by foreground ionized material.

Since intracloud extinction will affect the three observing bands differently, it is of interest to consider the three source populations and their locations with respect to the molecular cloud.  As a proxy, we show the contours of the submillimeter dust emission, as observed with SCUBA at 850~$\mu$m and reported by \citet{dif08}. The area to the west of the cluster that is clearly devoid of sources in all bands (see Figures~\ref{fig_innerarea_XIR} and \ref{fig_spatialdist}) is indeed in an area where more dust emission occurs when compared to the area east of the cluster. Since, however, centimeter wavelength sources are barely affected by extinction, contrary to near-infrared and X-ray sources, this suggests that the lower number count of radio sources is intrinsic to the cluster.

\begin{figure}
\includegraphics[width=\linewidth, bb=8 157 563 610]{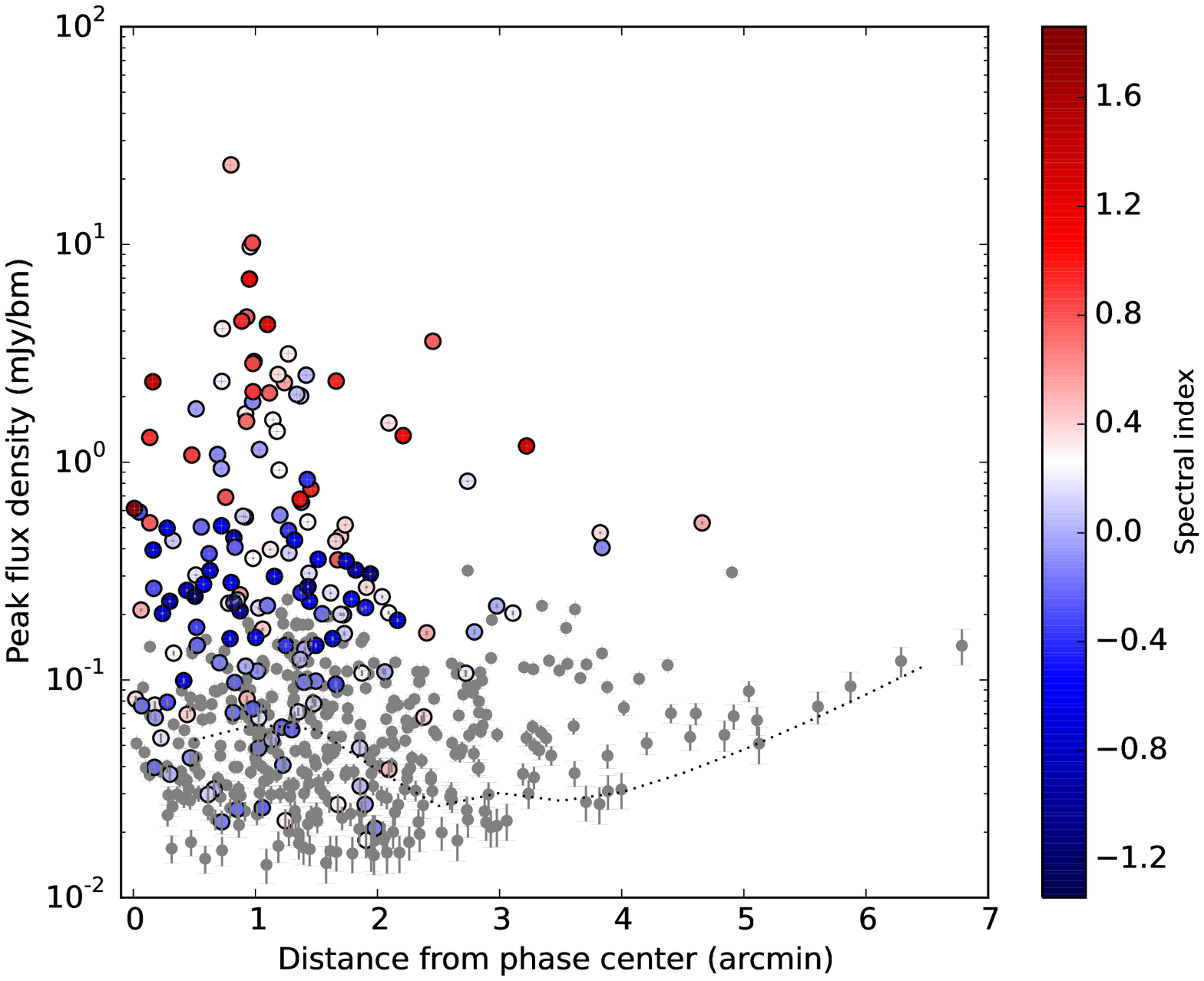}
\caption{Distribution of peak flux densities as a function of distance from the phase center. Sources with spectral indices that were determined with errors $\leq0.5$ are marked in color. The dotted line shows the approximate 5$\sigma$ limit, as derived in Figure~\ref{fig_rms_radial}.\label{fig_fluxdistr}}
\end{figure}

\subsection{Source properties}

In a significant change when compared to quasi-monochromatic observations with the pre-upgrade VLA, the primary beam correction for wide-band data has become more complicated \citep{bha13}. For our observation, the observing frequency, and thus the size of the primary beam, changes by a factor of $\sim$2.  To approximately correct flux densities for the decrease of sensitivity with increasing distance from the phase center, we have used the preliminary CASA task `widebandpbcor.' As noted above, this task also computes approximate spectral indices.  Particularly if there is an interest in sources with negative spectral indices, like here, where we use this as a criterion for nonthermal emission, the correction for the wideband primary beam response is very important.  In uncorrected data, the smaller primary beam size at higher frequencies would bias results toward negative spectral indices.

\begin{figure}
\includegraphics[width=1.1\linewidth]{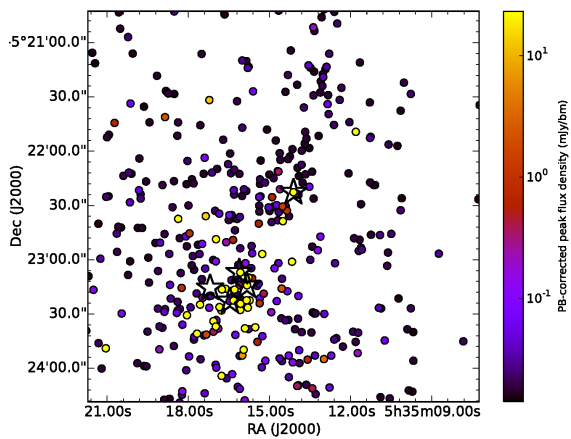}
\caption{Spatial distribution of peak flux densities in the inner area of the cluster. Star symbols denote the locations of the Trapezium stars and BN. \label{fig_inner_flux}}
\end{figure}

The radial distribution of peak flux densities is shown in Figure~\ref{fig_fluxdistr}.  A close-up view of the distribution of the peak flux densities in the inner cluster is shown in Figure~\ref{fig_inner_flux}. The brightest source detected is $\theta^1$~Ori~A with a peak flux density of 23.2~mJy\,bm$^{-1}$ (source 254). Without a primary-beam correction, this source is about 1800 times brighter than the weakest source reported as part of this list.  The figure also shows that the brightest sources are clustered in the neighbourhood of the Trapezium.

\begin{figure}

\includegraphics[width=\linewidth]{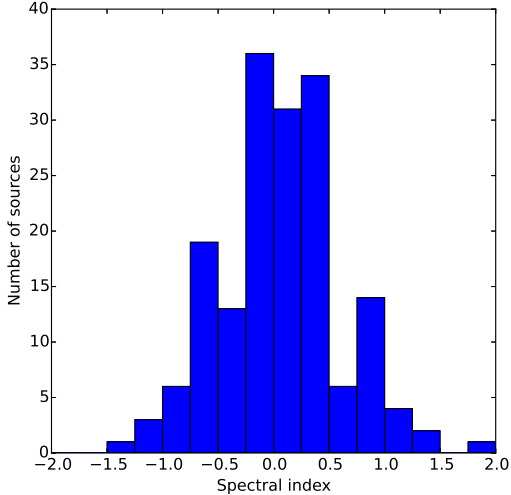}
\caption{Distribution of 170 radio spectral indices with errors $\leq0.5$.
\label{fig_si_hist}}
\end{figure}

Among the sources presented in this sample, 170 have spectral indices with errors $\leq0.5$, and these are marked in color in Figure~\ref{fig_fluxdistr}. Given the preliminary correction for the primary beam, we restrict the discussion of spectral indices to these sources.  The errors in these wideband spectral indices are calculated from the imaging residuals and are thus more realistic than errors based on the overall image noise alone. The brightest sources have positive spectral indices, while the sources with negative spectral indices are significantly weaker. This is plausible in a scenario where the sources with negative spectral indices are nonthermal coronal sources and the bright sources show thermal emission. Since the observed nonthermal radio emission is tied to the surface area of stellar coronae, the emitting volume is much more restricted than in the case of thermal emission from arbitrarily large amounts of ionized material (e.g., \citealp{gue02}). The distribution of the spectral indices in Figure~\ref{fig_si_hist} shows that the peak is close to zero, and with uncertainties of up to 0.5, at this point we cannot disentangle significantly positive and negative spectral indices for many of them.

\begin{figure}
\includegraphics[width=\linewidth,bb=-54 132 586 622]{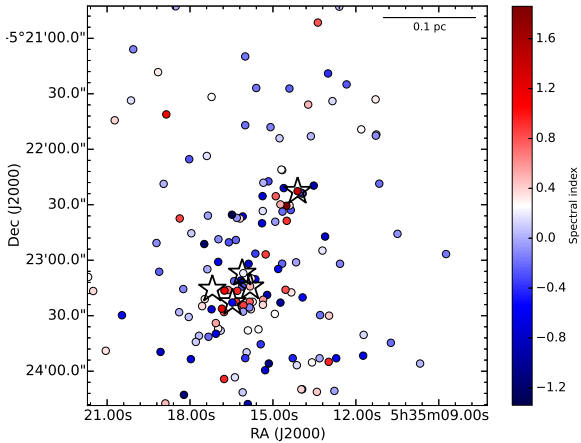}
\caption{Spatial distribution of spectral index determination in the inner cluster, only showing sources where the spectral index could be determined to an error of 0.5 or better. Star symbols denote the locations of the Trapezium stars and BN.\label{fig_inner_si}}
\end{figure}

For the inner area of the cluster, we show the spatial distribution of the spectral index measurements in Figure~\ref{fig_inner_si}. We again only plot sources where the uncertainty in spectral index is less than 0.5.  There is a cluster of sources with positive spectral indices, interpreted as thermal emission, close to the Trapezium. This corresponds to the main population of proplyds also shown in Figure~\ref{fig_inner}. However, there are also several sources with negative spectral indices in this inner area of the cluster, and there is thus no clear correlation between spectral indices and clustering.

A set of radio spectral indices for Orion sources was previously reported by \citet{kou14}. However, due to the lower sensitivity by more than a factor of 10, their spectral indices have larger errors with the `typical uncertainty' of 0.6, above our cutoff. While there is considerable disagreement in the reported spectral indices for individual sources, the numbers are mostly compatible within twice the reported errors, and as such the disagreements are not highly significant. A more detailed comparison would have to take into account differences in the calculation method of the spectral indices and may indicate significant variability in the spectral index for some sources.


\subsection{Background sources}

While we expect the source counts toward the center of the ONC to be dominated by sources related to the cluster, the presence of many sources with just a single-band detection in the inner $r=1\farcm6$ poses the question of contamination by extragalactic sources.  We only have limited means to assess membership with the data presented here, but single-band detections are a priori candidate extragalactic sources. A simple way to derive an upper limit to any population that may be unrelated to the cluster is to extrapolate the source counts in the outer beam to the inner area. Here, we use the annulus defined by the half-power beam widths in the two bands, as depicted in Figure~\ref{fig_psc}, where we neglect the South-Eastern quarter since it seems to contain a population of sources related to the ONC. Extrapolating the 21 sources contained in the remaining three quarters of this annulus and multiplying by the average primary beam correction factor for this annulus yields an estimate of 9 sources contained in an area with a radius of $1\farcm6$, i.e., the area depicted in the Venn diagram in Figure~\ref{fig_vennXIR}. This estimate indicates that 97\% of the compact radio sources, including the single-band detections, are related to the cluster.  This conclusion is supported by the fact that the single-band radio detections follow the same North-South elongation also seen in the full radio population and mentioned above.

Similarly, we note that \citet{get05b} have studied membership in the ONC based on the COUP X-ray data. In the inner area covered by the Venn diagram, they found only two candidate extragalactic sources among the full sample, indicating a contamination of the X-ray sample of only $\sim1$\%. We would expect a similar situation in the near-infrared given the excellent near-infrared--X-ray correspondence, and indeed \citet{mei16} find that the extinction levels in the area discussed here correspond to a negligible background contamination.

In summary, all three bands yield valuable information on the cluster, with only limited contamination from background sources.  While X-ray and near-infrared sources are very good tracers of the stellar cluster populations, the radio sources additionally trace features like jet shocks that are related to the cluster but not with a one-to-one correspondence with stellar sources.  Overall, the picture drawn for the inner cluster is the same, regardless of which wavelength we choose, but at the same time radio, X-ray, and near-infrared observations turn out to be highly complementary.


\subsection{The radio--X-ray connection}

\begin{figure}
\includegraphics*[width=\linewidth]{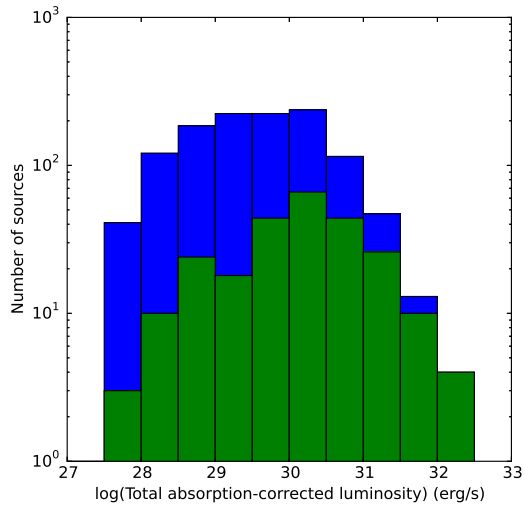}
\caption{Distribution of the absorption-corrected total X-ray luminosities of all X-ray sources in the field of view (blue) with the radio subset highlighted (green). \label{fig_Ltchist}}
\end{figure}

The comparison of X-ray and radio emission provides insight into the origin of the radio emission.  If there is an underlying connection between the X-ray emission and at least some of the radio emission (e.g., coronal activity that generates nonthermal radio and thermal X-ray emission), we would expect to see a correlation in the radio and X-ray luminosities. A total of 254 radio sources from our catalog have X-ray counterparts in the COUP survey. Compared to the entire set, the subset of X-ray sources with radio counterparts has no peculiar X-ray spectral properties (mainly, plasma temperature and foreground absorption). However, this subset generally selects X-ray sources with high luminosities. In Figure~\ref{fig_Ltchist}, we show histograms of the absorption-corrected X-ray luminosities (where available from COUP) of sources in the field of view and, highlighted in green, the subset of sources with radio counterparts. It becomes clear that all of the most luminous X-ray sources have radio counterparts, but the detection fraction rapidly drops toward lower X-ray luminosities. 

\begin{figure}
\includegraphics*[width=\linewidth]{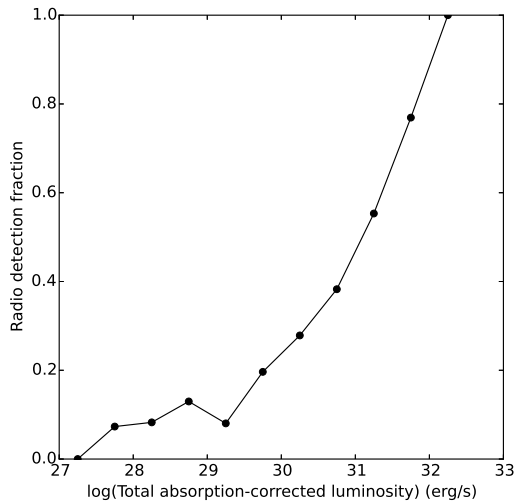}
\caption{Radio detection fraction as a function of absorption-corrected X-ray luminosity. \label{fig_ltc_fractions}}
\end{figure}

This effect has been discussed before (e.g., \citealp{fow13,riv15}), but it can be better quantified with the new sample. A plot of the corresponding detection fractions is shown in Figure~\ref{fig_ltc_fractions}. This correlation suggests that the fact that not all X-ray sources have radio counterparts is primarily a sensitivity effect. In our deep radio image, the detection rate drops to about 50\% for an absorption-corrected X-ray luminosity of log(L$_{\rm X}$)=31 erg\,s$^{-1}$.

Additionally, it is of particular interest to investigate the X-ray properties of candidate nonthermal radio sources. We here focus on the 17 radio sources with clear negative spectral indices (i.e. $>3\sigma$), ranging from $-0.21\pm0.02$ (source 27) to $-1.15\pm0.38$ (source 333).  Interestingly, not all of these sources have X-ray counterparts: only 10 out of these 17 sources have counterparts in the COUP survey.  Out of 45 sources with clear positive spectral indices (and presumably thermal radio emission), 37 have X-ray counterparts.

\begin{figure}
\includegraphics*[width=\linewidth,bb=20 157 561 607]{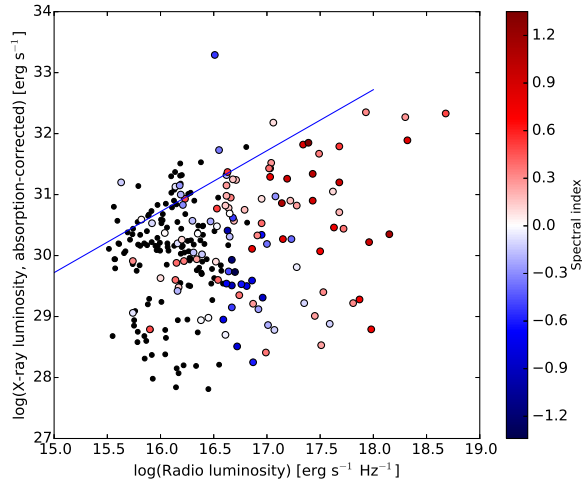}
\caption{X-ray and radio luminosities for sources that have absorption-corrected X-ray luminosities in COUP and a radio counterpart. Sources with radio spectral indices with nominal errors $\leq0.5$ are indicated in color. The blue line indicates the nominal G\"udel-Benz relation for active stars.\label{fig_gbplot3}}
\end{figure}

This finally leads us to consider a direct correlation of X-ray and radio luminosities, as observed in the G\"udel-Benz (GB) relation for active stars \citep{gue93,ben94}. For this purpose, we have converted the primary-beam--corrected radio flux densities to luminosities, assuming a common distance of 414~pc \citet{men07}. The resulting plot is displayed in Figure~\ref{fig_gbplot3}. This plot looks very similar to the version prepared by \citet{fow13} based on the archival VLA data of \citet{zap04} and COUP data.  With 254 sources, however, this plot contains more than five times as many points as the previous version.  Nevertheless, this plot still contains both thermal and nonthermal radio emitters, and the GB relation is only expected to apply to nonthermal radio emission. In terms of spectral types, this plot contains the full range from M to O stars. Some of the sources with the lowest X-ray and radio luminosities have been identified as M stars \citep{get05}, and both the most X-ray-luminous source as well as the most radio-luminous source in this plot are O stars. Other than that, there is no clear correlation with spectral type.

Figure~\ref{fig_gbplot3} shows that the X-ray and radio luminosities are not strongly correlated; there is a lot of scatter. The nominal GB relation serves as an approximate upper limit for the data. A priori, the problem is still that for a more reliable analysis, the nonthermal subset of the radio population would have to be identified. As a first step in this direction, we plot the 114 sources with a known X-ray luminosity where the radio spectral indices could be determined to an error of 0.5 or better (indicated in the color scale).  The radio and X-ray luminosities remain essentially uncorrelated, and there is also no apparent correlation with spectral indices. The sources with negative spectral indices are as uncorrelated as the full population. 

\subsection{The radio--infrared connection}

\begin{figure}
\includegraphics*[width=\linewidth]{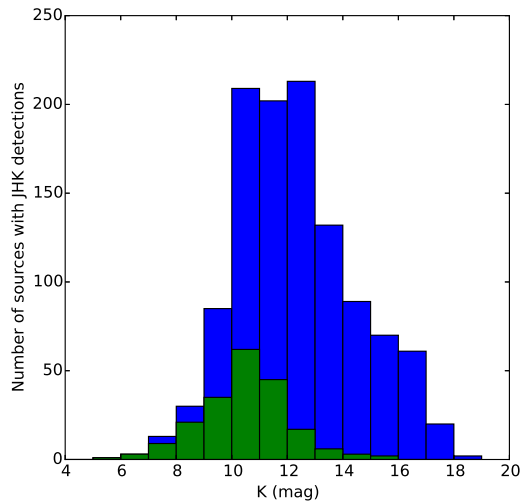}
\caption{Within $6\farcm7$ of the phase center, distribution of all VISION sources with {\it JHK} detections (blue), in comparison with those sources that have a radio counterparts (green). \label{fig_Khist}}
\end{figure}

Another interesting aspect in the characterization of the radio sources is the correspondence with infrared sources. Near-infrared observations provide a different census of the cluster, hampered only in some places by extinction and the presence of bright nebulosity, where the effective sensitivity is significantly reduced, as well as by the presence of foreground stars. In total, across the entire field, 258 radio sources in our catalog have near-infrared ($JHK_S$) counterparts in the VISION catalog \citep{mei16}. As mentioned above, these 258 sources constitute a population that is different from the 254 radio sources that have X-ray counterparts: a combined total of 214 radio sources have both X-ray and near-infrared counterparts.  A closer look at the properties of radio sources with near-infrared counterparts reveals that these sources have brighter near-infrared emission than the full VISION sample in the same field of view. This means that the radio detection fraction of infrared sources is at least partly a sensitivity effect. In Figure~\ref{fig_Khist}, we show a histogram visualizing this effect. The available near-infrared color information from the VISION catalog does not show that these colors are significantly different when comparing the near-infrared sources with and without radio counterparts.

\subsection{Remarks on individual sources}

\paragraph*{Trapezium stars}
Like \citet{kou14}, we detect radio emission from three of the four Trapezium stars, except $\theta^1$ Ori D. Like before, significantly variable radio emission is detected in the vicinity of $\theta^1$ Ori A, even if this radio source (GMR~12) is in fact a companion of $\theta^1$ Ori A \citep{pet98,pet07}. Additionally, radio emission is detected toward one of the two X-ray--emitting components of $\theta^1$ Ori B.  Interestingly, \citet{kou14} report a detection very close to the position of $\theta^1$ Ori C, with a flux density of $1.64\pm0.39$~mJy (at 4.5~GHz), while we only detect the source at a peak flux density (full band) of $0.16\pm0.03$~mJy, possibly a sign of significant variability.

\begin{figure}
\includegraphics*[width=\linewidth]{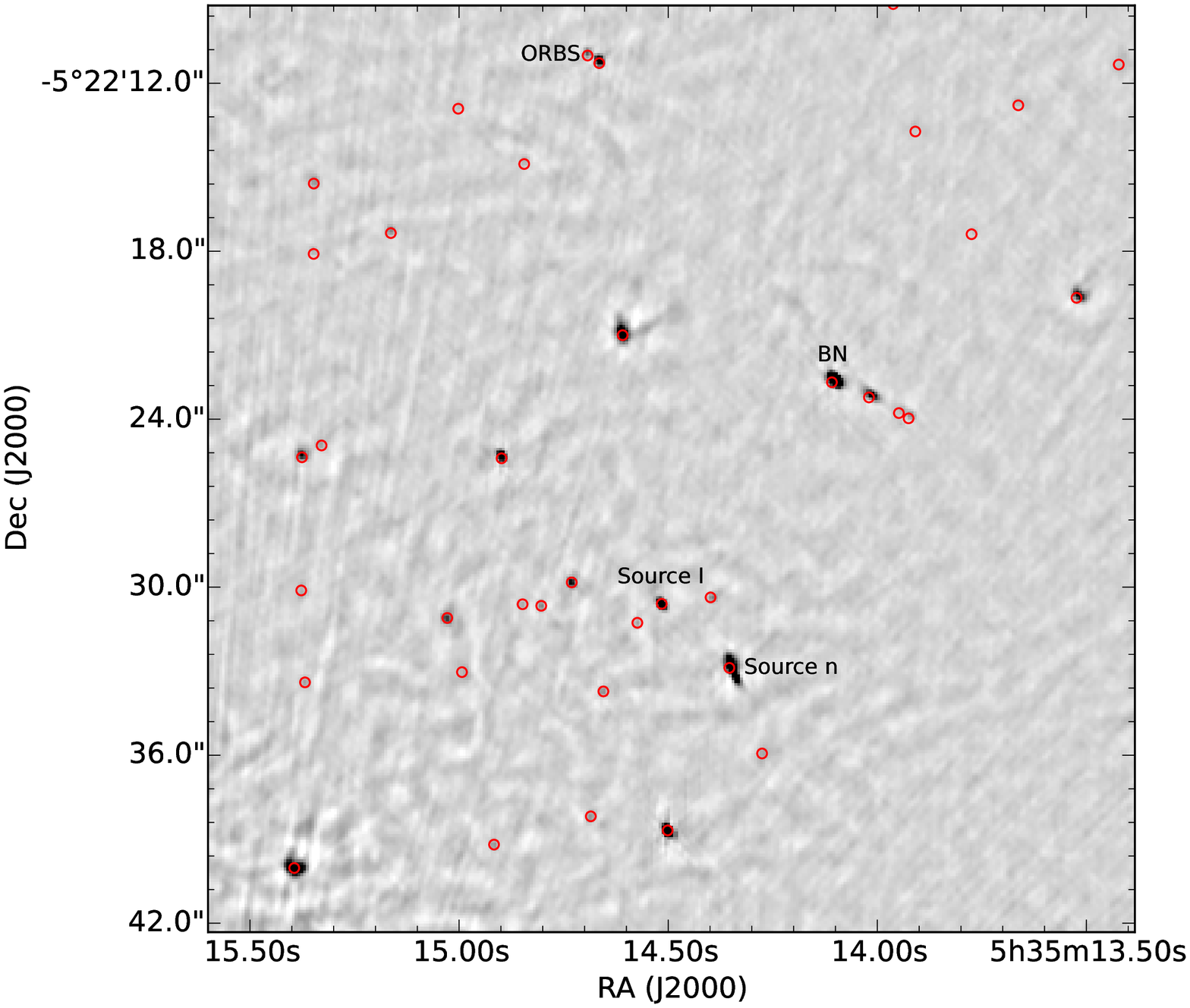}
\caption{The surroundings of BN and source I in our final continuum image ($(u,v)>100$k$\lambda$). The red circles mark objects from the compact source catalog. Next to source I and BN, also sources $n$ and ORBS, the radio flare source from \citet{for08} are labeled. \label{fig_bnkl}}
\end{figure}

\paragraph*{Source I} \cite{men95} identified source $I$ as a very luminous radio source, powering the Kleinmann-Low infrared nebula. As shown in Figure~\ref{fig_bnkl}, in the immediate surroundings of sources $I$ and $n$ several new faint sources were detected.  We detect source~I as catalog number 185 with a spectral index of 1.86$\pm$0.26, indicating thermal emission. The position is 43~mas away from the position reported by \citet{gom08}. Within the errors, the derived proper motion in RA and Dec is compatible with the values obtained by \citet{gom08}. Note that source $I$ is an example of a source that is likely associated with the cluster, yet does not have an X-ray or a near-infrared counterpart, presumably due to high extinction.

\paragraph*{Flare sources} The spectacular flare source ORBS, reported by \citet{for08}, is detected in our survey, but it does not show variability anywhere near the levels reported earlier.  This is only the second reported detection of this source outside of a flare.  It is clearly part of an apparent binary with two components that are $0\farcs5$ apart, with both components listed in our catalog (sources 198 and 201, see Figure~\ref{fig_bnkl}). Source 198 is the counterpart of ORBS, and it is listed with a peak flux density of $43.6\pm2.7$~$\mu$Jy\,bm$^{-1}$. It is thus brighter than the second component, source 201, listed at $13.3\pm2.7$~$\mu$Jy\,bm$^{-1}$. We also detect GMR~A, one of the longest-known radio sources in the ONC. A radio flare toward GMR~A was reported by \citet{bow03}. The variable source OHC-E, reported for the first time by \citet{riv15} is detected in all epochs.

\paragraph*{Proplyds} To find radio counterparts to known proplyds, we compare our deep radio source list with the catalog of \citet{ric08}. This catalog contains 197 circumstellar disks within $6\farcm7$ of the phase center in our data. Finding counterparts of resolved sources is not entirely obvious from catalog correlation, but we find 124 matches with a search radius of $1''$ and 130 matches with a search radius of $5''$.  About two thirds of the catalogued proplyds thus have compact radio counterparts. A more detailed look at some nondetections reveals that indeed no radio emission, not even from extended structures, is detected. As shown in Figure~\ref{fig_inner}, the radio emission of some of the proplyds is resolved in these observations, revealing the full proplyd geometry in great detail.

\paragraph*{Brown dwarfs} The fact that there is essentially no population of faint near-infrared sources with radio counterparts suggests in particular that the number of radio-detected brown dwarfs is very limited.  In our deep radio observations, one could have expected to detect radio emission of brown dwarfs (e.g., \citealp{ber01}). However, we have checked several catalogs of substellar objects (defined here as $>$M6.5) in the ONC, and we here report the detection of radio counterparts for three such objects.  We detect the M7.25 star 0990-3386, as reported by \citet{ing14}. It shows weak radio emission of (28$\pm$3)~$\mu$Jy\,bm$^{-1}$ (source 24). This star is also a known weak X-ray source.  The ratio of the radio and X-ray luminosities, using the absorption-corrected X-ray luminosity from COUP thus is log($L_R/L_X$) = --13.2\,Hz$^{-1}$, which falls onto the relation reported by \citet{wil14}. We also report the detection of a radio counterpart to [HC2000] 114, an M7 dwarf \citep{sle04}.  This is source 535 in our catalog, with a flux density of (54$\pm$5)~$\mu$Jy\,bm$^{-1}$, and here we find log($L_R/L_X$) = --12.3\,Hz$^{-1}$. The X-ray detection of this brown dwarf was also discussed by \citet{pre05bd}.  Finally, we report the detection of a radio counterpart to [HC2000] 764, an M7.5 dwarf. This infrared spectral classification is listed as less certain than $\pm1.5$ spectral type subclasses, and the visible-light classification is listed as K8--M3 \citep{sle04}, so this may not be a substellar object.  It is the brightest radio detection among the (candidate) brown dwarfs discussed here, at (155$\pm$3)~$\mu$Jy\,bm$^{-1}$ (source 379), and it does not have an X-ray counterpart.
	
\section{Summary and conclusions\label{sec:sum}}

We present a deep centimeter radio census of the Orion Nebula Cluster, based on a single 30~h pointing with the upgraded VLA in its high-resolution A-configuration and largely simultaneously with {\it Chandra} X-ray observations.  With the increase in sensitivity, the VLA is tracing a variety of complex structures, and we here focus on providing a catalog of 556 compact radio sources in the innermost area of the ONC.  The most extensive catalog of previously known sources, based on VLA observations prior to the VLA upgrade, listed 77 sources for the same area. We do not detect four bright radio sources from a survey by \citet{kou14}, possibly indicating strong variability for these sources. 

In a first step toward identifying radio emission mechanisms, we report centimeter radio spectral indices for 170 sources where the spectral index uncertainty is $\le0.5$. Both positive and negative spectral indices are found for ONC sources, tracing both thermal and nonthermal emission. Our spectral indices generally differ from those reported by \citet{kou14}, but at low significance. This may, however, be a signature of spectral index variability. Differences in the procedures used to calculate spectral indices may also play a role, and we compare our results based on MFS with manually calculated spectral indices, finding good general agreement for a subset of sources where a comparison is possible.

When comparing this new radio compact source catalog to the COUP deep X-ray census as well as to a new near-infrared survey of Orion A we find that their cumulative radial distributions are almost identical, in absolute terms even, which could be interpreted to mean that these three bands seem to effectively trace the cluster about equally well in the innermost region of $\sim1\farcm6$.  However, while these distributions trace each other exceedingly well, the radio, X-ray, and infrared source populations are quite distinct, with only a limited overlap. Yet, the spatial distributions of X-ray and near-infrared sources appear very similar also in 2D, but the radio source population is more concentrated toward both the Trapezium and the BN/KL region. Looking at individual sources, there is a clearer correlation between X-ray and near-infrared emission than between radio and either X-ray or near-infrared emission. In summary, while the radio sources appear to be overwhelmingly related to the cluster, not all of them are stellar but instead also include compact emission from ionized material (e.g., from jet shocks).

A comparison of the radio catalog with the X-ray and near-infrared catalogs shows that in spite of the considerable increase in sensitivity, the radio observations are still sensitivity-limited when compared to the other two bands.  Among the X-ray sources with radio counterparts and known absorption-corrected X-ray luminosities, the radio detection fraction approaches 100\% only for the most X-ray--luminous objects, before falling quickly toward lower luminosities, reaching 50\% at an absorption-corrected X-ray luminosity of log(L$_{\rm X}$)=31 erg\,s$^{-1}$. This completeness analysis can be used to plan radio observations of other young clusters of X-ray--emitting YSOs. Similar to previous studies, preliminary results additionally show that there is no clear correlation of X-ray and radio luminosities, but a conclusive analysis will require carefully disentangling nonthermal and thermal emission first.  Also for the near-infrared sources with radio counterparts, the radio detection fraction is highest for the brightest near-infrared sources, again underlining the relative sensitivity limitations.

This radio source catalog provides a target list for our ongoing study of correlated X-ray and radio variability at unprecedented time resolution and sensitivity.  Additionally, we are working on a detailed study of thermal and nonthermal sources in the area, involving more detailed wideband primary beam correction. Furthermore, we have obtained astrometric follow-up observations with the Very Long Baseline Array. Finally, this ONC radio dataset of unprecedented sensitivity also offers an excellent starting point not only for studies of compact sources, but also for studies of extended features like proplyds, bow shocks and filamentary structures in the Orion Nebula, particularly when it comes to astrometric studies.

\acknowledgments 
Support for this work was provided by the National Aeronautics and Space Administration through Chandra Award Number GO2-13019X issued by the Chandra X-ray Observatory Center, which is operated by the Smithsonian Astrophysical Observatory for and on behalf of the National Aeronautics Space Administration under contract NAS8-03060. JF acknowledges informative discussions with Eric Feigelson and Kosta Getman. JF and VMR acknowledge the hospitality of NRAO Socorro during an extended data reduction visit. VMR is funded by the Italian Ministero dell'Istruzione, Universit\`a e
Ricerca through the grant Progetti Premiali 2012 -- iALMA. This research made use of APLpy, an open-source plotting package for Python hosted at http://aplpy.github.com; Astropy, a community-developed core Python package for Astronomy \citep{ast13}; matplotlib, a Python library for publication quality graphics \citep{hun07}. 



{\it Facilities:} \facility{NRAO (VLA)} \facility{CXO (ACIS)}.






\bibliographystyle{apj.bst}
\bibliography{orion}

\clearpage

\begin{deluxetable}{rrrrrrrrr}
\tabletypesize{\scriptsize}
\rotate
\tablecaption{Orion Nebula Cluster Radio Compact Source Catalog\label{tbl-1}}
\tablewidth{0pt}
\tablehead{
\colhead{Nr.} & \colhead{RA} & \colhead{Dec} & \colhead{Peak\tablenotemark{b} (mJy)} & \colhead{Sp. index} & \colhead{IDs\tablenotemark{c}} & \colhead{COUP} & \colhead{Ltc\tablenotemark{d}} & \colhead{VISION}}
\startdata
1  & 05 34 47.99621 $\pm$ 0.0039080 & -05 20 54.1061 $\pm$ 0.022215 & 0.144 $\pm$ 0.026 & \nodata       & K	 & \nodata  & \nodata & \nodata 	 \\
2  & 05 34 55.98002 $\pm$ 0.0002125 & -05 23 12.8732 $\pm$ 0.001889 & 0.526 $\pm$ 0.008 & 0.52$\pm$0.29 & K	 &  107     & 31.43   & 05345597-0523130 \\ 
3  & 05 34 56.68496 $\pm$ 0.0006779 & -05 20 33.3647 $\pm$ 0.020391 & 0.056 $\pm$ 0.009 & \nodata       & \nodata & \nodata  & \nodata & \nodata 	 \\
4  & 05 35 00.11291 $\pm$ 0.0013979 & -05 23 01.9841 $\pm$ 0.017545 & 0.037 $\pm$ 0.005 & \nodata       & \nodata &  141     & 30.65   & 05350011-0523019 \\ 
5  & 05 35 02.00589 $\pm$ 0.0003720 & -05 20 55.0863 $\pm$ 0.005465 & 0.110 $\pm$ 0.005 & \nodata       & \nodata &  172     & 30.49   & 05350200-0520551 \\ 
6  & 05 35 02.07950 $\pm$ 0.0017977 & -05 26 36.0192 $\pm$ 0.027732 & 0.051 $\pm$ 0.010 & \nodata       & \nodata &  173     & 30.58   & 05350208-0526363 \\ 
7  & 05 35 03.63038 $\pm$ 0.0004761 & -05 20 02.2630 $\pm$ 0.006567 & 0.102 $\pm$ 0.005 & \nodata       & \nodata & \nodata  & \nodata & \nodata 	 \\
8  & 05 35 04.85920 $\pm$ 0.0000088 & -05 23 02.6173 $\pm$ 0.000129 & 3.592 $\pm$ 0.004 & 0.75$\pm$0.06 & Z, K	 &  229     & 29.28   & \nodata 	 \\
9  & 05 35 04.95277 $\pm$ 0.0009842 & -05 21 09.2055 $\pm$ 0.017297 & 0.025 $\pm$ 0.004 & \nodata       & \nodata &  230     & 30.26   & 05350495-0521092 \\ 
10 & 05 35 06.28447 $\pm$ 0.0001158 & -05 22 02.6438 $\pm$ 0.001907 & 0.203 $\pm$ 0.003 & 0.27$\pm$0.10 & K	 &  262     & 31.15   & 05350628-0522027 \\ 
\enddata
\tablecomments{Table \ref{tbl-1} is published in its entirety in the 
electronic edition of the {\it Astrophysical Journal}.  A portion is 
shown here for guidance regarding its form and content.}
\tablenotetext{a}{Peak flux density from Gaussian fit, corrected for primary beam response using CASA task widebandpbcor.}
\tablenotetext{b}{Spectral index, corrected for primary-beam response (see text), if error $\leq0.5$}
\tablenotetext{c}{Previous radio designations in \citet{zap04} or \citet{kou14}, designated as Z an K.}
\tablenotetext{d}{Total absorption-corrected X-ray luminosity from COUP \citep{get05}, where available.}
\end{deluxetable}

\end{document}